\documentclass[fleqn,12pt]{article} 
\usepackage{graphicx}
\usepackage{epsfig} 
\usepackage{wrapfig}
\usepackage{amsmath}
\begin{document}
\noindent
{\Large \bf On the Feynman-alpha formula for fast neutrons}
\newline
\newline
\begin{center}
{Johan Anderson\footnote{johan@nephy.chalmers.se}, L\'en\'ard P\'al$^{2}$ and Imre P\'{a}zsit$^{1}$ }
\end{center}
\begin{center} 
$^1$ Department of Nuclear Engineering, Chalmers University of Technology, SE-41296, G\"{o}teborg, Sweden \\
\end{center}
\begin{center}
$^{2}$ KFKI Atomic Energy Research Institute H-1525 Budapest 114, POB 49, Hungary
\end{center}

\begin{abstract}
\noindent
In this contribution, a stochastic theory for a branching process in a neutron population with two energy levels is investigated. In particular, a variance to mean or Feynman-alpha formula is derived in this generalized scenario using the Kolmogorov forward or master equation theory for the probabilities in a system with a compound Poisson source.
\end{abstract}

\section{Introduction}
There exist several relatively new applications where the energy distribution of the neutrons plays a significant role. One particular case is a method used in nuclear safeguards, namely the stochastic generalization of the so-called differential die-away analysis (DDAA) \cite{kunz1982, croft2003, jordan2007, jordan2008}. Traditionally, the DDAA method was used as a deterministic method of detecting fissile material embedded in moderating surroundings using a pulsed source. The newly explored method, called differential die-away self-interrogation (DDSI) utilizes the inherent spontaneous neutron emission of the sample \cite{menlove09}. In the DDSI method the temporal decay of the correlations as a function of the time delay between two detections of fast neutrons is used. This corresponds to a Rossi-alpha measurement with two energy groups. Furthermore, in recent pulsed experiments measuring the reactivity in fast cores of accelerator driven sub-critical systems it is found that two exponentials appear, indicating that the temporal behavior of the fast and thermal neutrons is separated in fast reflected cores. This leads to the fact that a two group versions of the Feynman and Rossi-alpha formulas are needed \cite{a1}-\cite{a2}.

In this contribution, a stochastic theory for a branching process in a neutron population with two energy levels is investigated based on the previous results in Refs~\cite{a1}-~\cite{a2}. In particular a variance to mean or Feynman-alpha formula is derived in this generalized scenario using the master equation or Kolmogorov forward approach. The model includes absorbtions, down scattering from fast to thermal neutrons, thermal fissions, detections and an external source of fast neutrons. Higher moments will also be discussed as well as specific applications to areas within safeguards research as well as specific applications to areas within safeguards research.

\section{The variance to mean via the forward Kolmogorov approach}
In this section we will describe the two particle type system by using the Kolmogorov forward approach. We will include a compound Poisson source of fast neutrons described by the source strength $S_1$ which releases $n$ particles with probability $p_q(n)$ at an emission event (i.e. spontaneous fission). The source is assumed to be switched on at time $t=t_0$, although dependence on $t_0$ will not be denoted. The effects of detecting particles will also be included, denoted by the intensity $\lambda_d$. We will start by giving the differential equation for the probability $P(N_1, N_2, Z_1,t)$ for $N_1$ fast, $N_2$ thermal neutrons at time $t$ and $Z_1$ detected fast particles in the interval $(0,t)$. We have summed all mutually exclusive events during an infinitesimally small time interval $dt$ and find for probabilities,
\begin{eqnarray}\label{eq:1.1}
\frac{\partial P(N_1,N_2,Z_1,t)}{\partial t} & = & - (\lambda_1 N_1 + \lambda_2 N_2 + S_1)P(N_1, N_2, Z_1,t) \nonumber \\
& + & \lambda_{1a} (N_1 + 1) P(N_1 + 1, N_2, Z_1,t)\nonumber \\
& + & \lambda_{2a} (N_2 + 1) P(N_1, N_2 + 1, Z_1,t) \nonumber \\
& + & \lambda_R (N_1 + 1) P(N_1 + 1, N_2 - 1, Z_1, t)\nonumber \\
& + & \lambda_{2f} (N_2 + 1) \sum_k^{N_1} f(k) P(N_1 - k, N_2 + 1, Z_1,t)\nonumber \\
& + & \lambda_d (N_1 + 1) P(N_1 + 1, N_2, Z_1 - 1,t) \nonumber \\
& + & S_1 \sum_n^{N_1} p_q(n) P(N_1 - n, N_2, Z_1,t).
\end{eqnarray}
Here, $\lambda_1$ and $\lambda_2$ are the decay constants (total reaction intensities) for fast and thermal neutrons whereas $\lambda_{1a}$, $\lambda_{2a}$ are the absorbtion intensities of fast and thermal particles, respectively. The removal of fast particles into the thermal group is described by $\lambda_R$ while fission resulting from the thermal particles happens with the intensity of $\lambda_{2f}$. The intensities are related through
\begin{eqnarray}\label{eq:1.2}
\lambda_1 = \lambda_{1a} + \lambda_R + \lambda_d,
\end{eqnarray}
and
\begin{eqnarray}\label{eq:1.3}
\lambda_2 = \lambda_{2a} + \lambda_{2f}.
\end{eqnarray}
We will now solve this differential equation by using the generating function of the form
\begin{eqnarray}\label{eq:1.4}
G(X,Y,Z,t) & = & \sum_{N_1} \sum_{N_2} \sum_{Z_1} X^{N_1} Y^{N_2} Z^{Z_1} P(N_1,N_2,Z_1,t),
\end{eqnarray}
and describe the process in the time evolution of the generating function as,
\begin{eqnarray}\label{eq:1.5}
\frac{\partial G}{\partial t} & = & (\lambda_{1a}  + \lambda_R Y + \lambda_d Z - \lambda_1 X) \frac{\partial G}{\partial X} + (\lambda_{2a}  + \lambda_{2f} \nu (X) - \lambda_2 Y) \frac{\partial G}{\partial Y} \nonumber \\
& + & S_1 (r(X) - 1) G,
\end{eqnarray}
where
\begin{eqnarray}
\nu (X) & = & \sum_k f_k X^k, \label{eq:1.61} \;\;\;\;\;\;\;\;\;\;  \mbox{and} \\
r(X) & = & \sum_n p_q(n) X^n. \label{eq:1.62}
\end{eqnarray}
Here, $f_k$ is the probability of having exactly $k$ neutrons produced in a fission event. Differentiation of equation (\ref{eq:1.5}) with respect to ($X,Y,Z$) and then letting ($X = Y = Z = 1$) yields differential equations for the expectations as,
\begin{eqnarray}
\frac{\partial }{\partial t} \langle N_1 \rangle & = & -\lambda_1 \langle N_1 \rangle + \lambda_{2f} \nu_1 \langle N_2 \rangle + S_1 r_1, \label{eq:1.7}\\
\frac{\partial }{\partial t} \langle N_2 \rangle & = & -\lambda_2 \langle N_2 \rangle + \lambda_{R} \langle N_1 \rangle , \label{eq:1.8} \\
\frac{\partial }{\partial t} \langle Z_1 \rangle & = & \lambda_d \langle N_1 \rangle. \label{eq:1.9}
\end{eqnarray}
Here we have used the definition of the derivatives on the equations (\ref{eq:1.61}) and (\ref{eq:1.62}) as $\nu_1 = d q/dX|_{X=1}$ and $r_1 = d h/dX|_{X=1}$. We note that due to the source term with intensity $S_1$ the dynamical system consisting of equations (\ref{eq:1.7}) - (\ref{eq:1.8}) will reach a steady state ($\frac{\partial}{\partial t} \rightarrow 0$) and we find the stationary solution,
\begin{eqnarray}
\langle N_1 \rangle & = & \bar{N}_1 = \frac{\lambda_2 S_1 r_1}{\lambda_1 \lambda_2 - \nu_1 \lambda_R \lambda_{2f}} =  \frac{\lambda_2 S_1 r_1}{\omega_1 \omega_2}, \label{eq:1.10}\\
\langle N_2 \rangle & = & \bar{N}_2 = \frac{\lambda_R S_1 r_1}{\omega_1 \omega_2}, \label{eq:1.11} \\
\langle Z_1 \rangle & = & \epsilon \lambda_{2f} \bar{N}_1 t, \label{eq:1.12}
\end{eqnarray}
where $\epsilon = \lambda_d/\lambda_{2f}$ and we have used the additional definitions $\omega_1$ and $\omega_2$,
\begin{eqnarray}
-\omega_1 & = & -\frac{1}{2}(\lambda_1 + \lambda_2) + \frac{1}{2} \sqrt{(\lambda_1 - \lambda_2)^2 + 4 \lambda_1  \lambda_2 \nu_{eff}} ,\label{eq:1.13} \\
-\omega_2 & = & -\frac{1}{2}(\lambda_1 + \lambda_2) - \frac{1}{2} \sqrt{(\lambda_1 - \lambda_2)^2 + 4 \lambda_1  \lambda_2 \nu_{eff}},\label{eq:1.14} \\
\nu_{eff} & = & \nu_1 \frac{\lambda_R \lambda_{2f}}{\lambda_1 \lambda_2}. \label{eq:1.15}
\end{eqnarray}
The expectation of the detections is found by integrating equation (10) and we note that the number of detections increases linearly with time. In order to find the variance of the detector counts we need to determine the second factorial moment by yet another differentiation with respect to ($X,Y,Z$) and then letting ($X = Y = Z = 1$). The variance of the detector counts can be determined through the relation $\sigma_Z^2 = \langle Z_1 \rangle + \mu_{ZZ}$ where the modified variance $\mu_{ZZ}$ is defined as $\mu_{ZZ} = \langle Z(Z-1) \rangle - \langle Z\rangle^2 = \sigma_{ZZ}^2 - \langle Z\rangle$ while in general we have $\mu_{X Y} = \langle X Y\rangle - \langle X \rangle \langle Y \rangle$. The differentiation procedure gives a system of six dynamical equations of the modified second moments as
\begin{eqnarray}
\frac{\partial}{\partial t} \mu_{X X} & = & - 2 \lambda_1 \mu_{X X} + 2 \nu_1 \lambda_{2 f} \mu_{X Y}  + \nu_2 \lambda_{2f} \bar{N}_2 + S_1 r_2, \label{eq:1.16} \\
\frac{\partial}{\partial t} \mu_{X Y} & = & -(\lambda_1 + \lambda_2) \mu_{X Y} + \lambda_R \mu_{X X} + \nu_1 \lambda_{2 f} \mu_{Y Y}, \label{eq:1.17}\\
\frac{\partial}{\partial t} \mu_{Y Y} & = & - 2 \lambda_2 \mu_{Y Y} + 2 \lambda_R
 \mu_{X Y},\label{eq:1.18} \\
\frac{\partial}{\partial t} \mu_{Z X} & = & - \lambda_1 \mu_{Z X} + \nu_1 \lambda_{2f} \mu_{Z Y} + \lambda_d \mu_{X X},\label{eq:1.19} \\
\frac{\partial}{\partial t} \mu_{Z Y} & = & - \lambda_2 \mu_{Z Y} + \lambda_R \mu_{Z X}  + \lambda_d  \mu_{X Y}, \label{eq:1.20} \\
\frac{\partial}{\partial t} \mu_{ZZ} & = & 2\epsilon \lambda_{2f} \mu_{X Z}, \label{eq:1.21}
\end{eqnarray}
where we have used the additional notations $\nu_2 = d^2 q/dX^2|_{X=1}$ and $r_2 = d^2 h/dX^2|_{X=1}$. Although, the system of equations (\ref{eq:1.16}) - (\ref{eq:1.21}) is rather complicated and an analytical solution would be hard to find, we note that in stationary state the system breaks down into two systems independent of each other where the moments  $\mu_{X X} = \bar{\mu}_{XX} $, $\mu_{X Y} = \bar{\mu}_{XY} $ and $\mu_{Y Y} = \bar{\mu}_{YY} $ are constants. However, the equations describing detected particles need to be solved by e.g. Laplace transforms of (\ref{eq:1.19}) and (\ref{eq:1.20}) whereas it is possible to find the sought moment $\mu_{Z Z} $ by integration by using equation (\ref{eq:1.21}). We find the constant 2nd modified moments as,
\begin{eqnarray}
\bar{\mu}_{XX} & = & \frac{(\lambda_2^2 + \omega_1 \omega_2)(\nu_2 \lambda_{2f} \bar{N}_2 + S_1 r_2)}{2(\lambda_1 + \lambda_2)\omega_1 \omega_2}, \label{eq:1.22}\\
\bar{\mu}_{XY} & = &  \frac{\lambda_2 \lambda_R (\nu_2 \lambda_{2f} \bar{N}_2 + S_1 r_2)}{2(\lambda_1 + \lambda_2)\omega_1 \omega_2}, \label{eq:1.23} \\
\bar{\mu}_{YY} & = & \frac{\lambda_R^2 (\nu_2 \lambda_{2f} \bar{N}_2 + S_1 r_2)}{2(\lambda_1 + \lambda_2)\omega_1 \omega_2}. \label{eq:1.24}
\end{eqnarray}
The objective now is to solve (\ref{eq:1.19}) and (\ref{eq:1.20}) by Laplace transform methods and we find the transformed identity as,
\begin{eqnarray}
\tilde{\mu}_{XZ} = \frac{\nu_1 \lambda_d \lambda_{2f} \bar{\mu}_{XY}}{s H(s)} + \frac{(s+\lambda_2)\lambda_d \bar{\mu}_{XX}}{s H(s)} \label{eq:1.25}
\end{eqnarray}
with
\begin{eqnarray}
H(s) = s^2 + (\omega_2 + \omega_1)s + \omega_1 \omega_2. \label{eq:1.26}
\end{eqnarray} 
Note that we have assumed that the initial values of the moments $\mu_{XZ}$ and $\mu_{YZ}$ were equal to zero at $t=0$ (at the start of the measurement), hence the roots of $H(s)$ determine the temporal behavior of the Feynman-alpha formula. Moreover, the solution has many similarities to that found in Ref. \cite{pazsit1999}. The variance $\sigma_{ZZ} = \langle Z \rangle + \mu_{ZZ}$ is now found by integration of (\ref{eq:1.21}) and after some algebra the Feynman-alpha formula can now be written in the form
\begin{eqnarray}
\frac{\sigma_{ZZ}(T)}{Z_1} = 1 + Y_1 (1 - \frac{1-e^{- \omega_1 T}}{\omega_1 T}) + Y_2 (1 - \frac{1-e^{- \omega_2 T}}{\omega_2 T}). \label{eq:1.27}
\end{eqnarray} 
Here, the complete expressions for $Y_1$ and $Y_2$ are quite lengthy. However, it turns out that the sum $Y_0 = Y_1 + Y_2$ takes a rather simple form that also determines the value of the Feynman-alpha for large measurement times $T \rightarrow \infty$ as,
\begin{eqnarray}
Y_0 =  Y_1 + Y_2 = q_2 \frac{\lambda_d \lambda_2 \lambda_R \lambda_{2f}}{\omega_1^2 \omega_2^2}. \label{eq:1.28}
\end{eqnarray} 
We will now turn our attention to some quantitative examples of the Feynman-alpha formula in the form of Equation (\ref{eq:1.27}). 
\section{Results and discussion}
The Feynman-alpha formula for a two particle type system found by using the Kolmogorov forward approach including a Poisson source and effects of detecting particles is shown in Figure 1 (A and B). We have used the parameters $\nu_1 = 3.0$, $\nu_2 = 5.0$, $S_1 = 1.0$, $r_1 = 1.0$, $r_2 = 0.0$, $\lambda_{2f} = 3/5$ and $\lambda_d = 0.1$. In Figure 1A $\lambda_R = 2/3$ whereas in Figure 1B $\lambda_1 = 1.0$ and $\lambda_2 = 2.0$. As expected, the curves grow according to Equation (\ref{eq:1.27}) exponentially to a maximum value determined by the constant $Y_0$. However, this value can signifcantly vary depending on the intensities ($\lambda_1$, $\lambda_2$, etc) involved the process. The increase of the curves is determined by two exponentials. Furthermore, unlike the case of the DDSI method, the presence of the two exponentials is not visible to the bare eye. In Figure 1A, it is seen that the ratio of the decay intensities for fast and thermal particles has a nontrivial effect on the maximum value by changing the ratio in the range ($0.25 - 4.0$). In Figure 1B, the effect of the thermalization process described by the intensity $\lambda_R$ on the results is illustrated. Increasing thermalization increases the asymptotic value of the Feynman-alpha. 
\begin{figure}[ht]
\begin{minipage}[b]{0.5\linewidth}
\centering
\includegraphics[width=7cm, height=6.5cm]{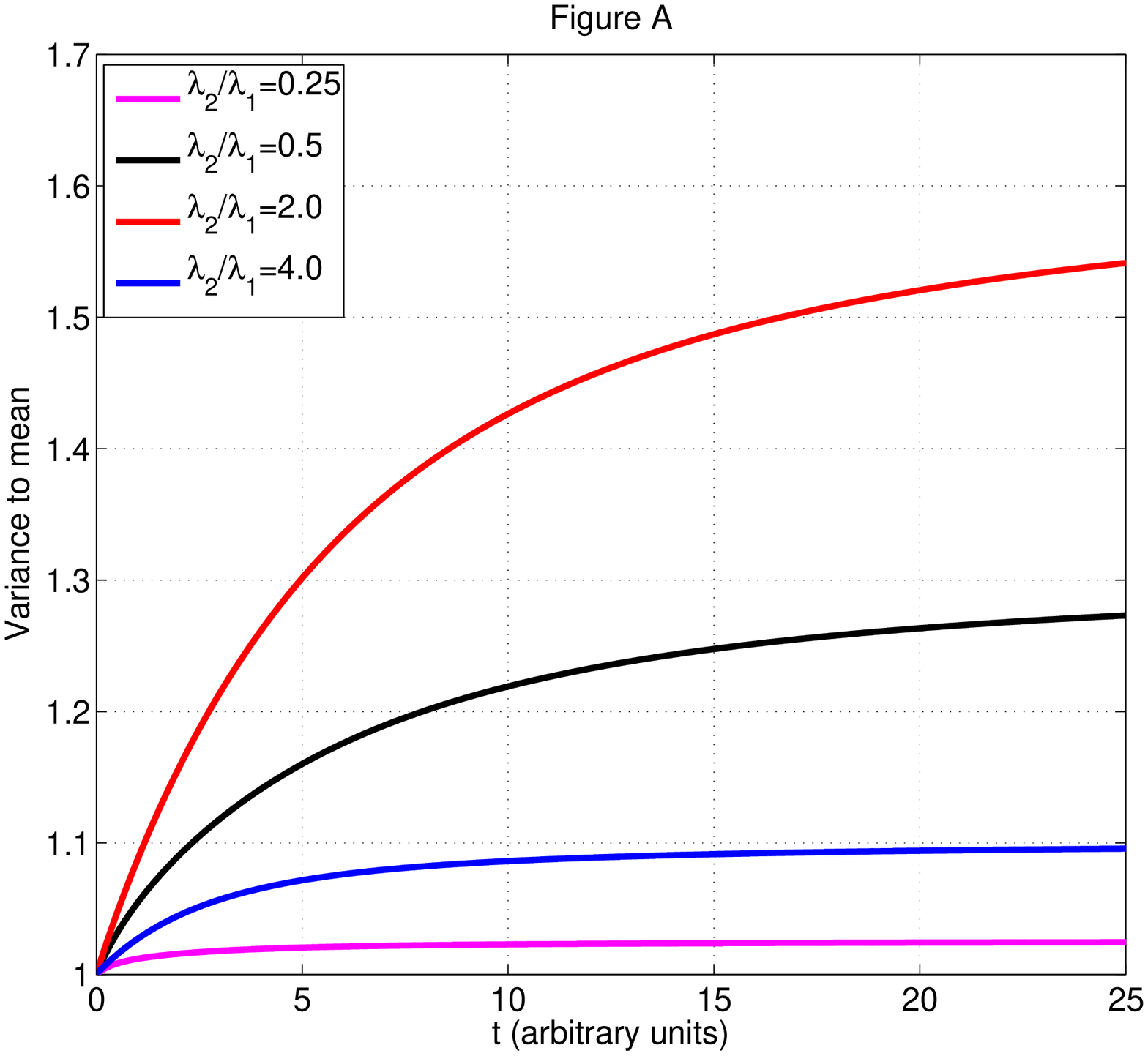}
\label{fig:figure1}
\end{minipage}
\hspace{0.5cm}
\begin{minipage}[b]{0.5\linewidth}
\centering
\includegraphics[width=7cm, height=6.5cm]{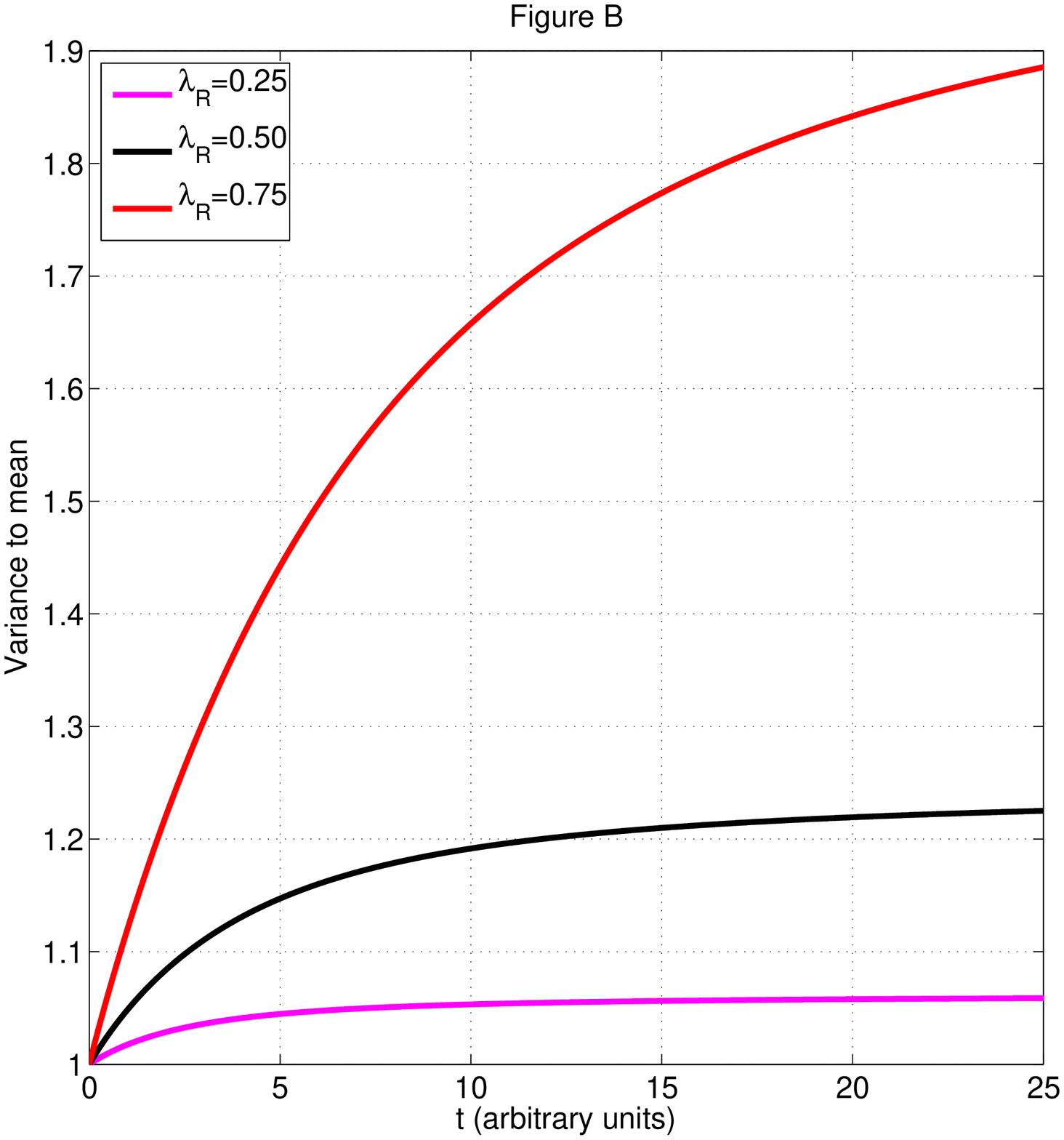}
\label{fig:figure2}
\end{minipage}
\caption{(A and B) The Feynman-alpha expression is shown for the parameters $\nu_1 = 3.0$, $\nu_2 = 5.0$, $S_1 = 1.0$, $r_1 = 1.0$, $r_2 = 0.0$, $\lambda_{2f} = 3/5$ and $\lambda_d = 0.1$. In Figure A, $\lambda_R = 2/3$ whereas in Figure B $\lambda_1 = 1.0$ and $\lambda_2 = 2.0$. }
\end{figure}
\section{Conclusions}
We have developed a forward Kolmogorov approach for the two group theory of the Feynman-alpha method, including a compound Poisson source and the detection process. The results agree with those calculated by the backward approach as reported in \cite{a1}. It is seen that, unlike in the DDSI method (i.e. the two-group version of the Rossi-alpha method), the presence of two exponents in the solution is not clearly visible. This means that detection of the presence of fissile material may not be as obvious as with the Rossi-alpha method. On the other hand, the determination of the exponents $\omega_1$ and $\omega_2$ by curve fitting could be more accurate in certain cases than with the DDSI method. However, the diagnostic value of the exponents in terms of determination of the sample parameters is not clear yet, and it requires further investigations, which will be reported in further work.
\section{Acknowledgements}
This work was supported by the Swedish Radiation Safety Authority (SSM).

\end{document}